 \documentclass[preprint,12pt]{elsarticle}

\usepackage{graphicx}
\usepackage{epsfig}
\usepackage{psfrag}
\usepackage{mathrsfs}
\usepackage{amssymb}
\usepackage{amsmath}
\usepackage{xcolor}
\usepackage[utf8]{inputenc}

\journal{Journal of Subatomic Particles and Cosmology}

\begin{document}

\begin{frontmatter}



\title{Harmonic Oscillator Representation of Scattering Theory in the Presence of Coulomb Potential}


\author{U. M. Yanikov\fnref{label1}\corref{cor1}}
  \ead{yanikov-u@yandex.ru}
\author{V. A. Kulikov\fnref{label1}}
  \ead{kulikov@nucl-th.sinp.msu.ru}
\author{A.~M.~Shirokov\fnref{label1}}
  \ead{shirokov@nucl-th.sinp.msu.ru}

\affiliation[label1]{organization={Skobeltsyn Institute of Nuclear Physics, Moscow State University},
            addressline={1-2 Leninskiye Gory Street}, 
            city={Moscow},
            postcode={119234},
            country={Russia}}

\cortext[cor1]{Corresponding author}

\begin{abstract}
    Considering the problem of scattering of 
    charged particles, 
we introduce a 
new approach of taking the Coulomb interaction into account within the HORSE formalism. Compared to the conventional HORSE approach 
{for uncharged particles}, 
{we add a diagonal Coulomb term to} the three-term recurrent 
relation for expansion coefficients in the asymptotic region. 
{The method simplifies calculations 
and demonstrates a good agreement with numerical solution.}
\end{abstract}


\end{frontmatter}

\section{Introduction}
\label{intro}

Modern calculations of 
bound states of light atomic nuclei are performed by \textit{ab initio} methods, i.\,e., by supercomputer calculations 
without relying on any model assumptions about the nuclear structure. The No-Core Shell Model (NCSM) \cite{BARRETT2013131} is one of the most advanced and promising approaches in this field. The 
{mainstream} direction in the development of 
{light nuclei} theory is the advancement of \textit{ab initio} methods 
describing nuclear resonance states and reactions.

The Harmonic Oscillator Representation of Scattering Equations (HORSE) formalism 
\cite{Yamani,osti_6451278,Filip,osti_6232444,Bang1999PmatrixAJ} is one of the approaches to studying continuous spectrum states which 
has been successfully applied to the studies of 
nuclear resonant 
states~
{\cite{PhysRevC.94.064320, PhysRevC.98.044624}}, 
photodesintegration~{\cite{PhysRevC.106.054608}},
three-body continuum within phenomenological cluster models~\cite{Mikh,Lurie} and various 
problems within the Resonating Group Model (RGM)~\cite{Arickx,Kato,Lashko}.
 Accounting for the Coulomb interaction between charged particles within the HORSE method is crucial but presents challenges due to the long-range nature of the Coulomb potential. Several approaches have been proposed to describe the Coulomb potential 
 impact~\cite{Bang1999PmatrixAJ, OKHRIMENKO1984121}. The method developed by the Kiev group \cite{OKHRIMENKO1984121} enables successful calculations within the 
 RGM~\cite{Okrimenko1987,Arickx,Kato,Lashko}. However, it requires computing sums of terms 
 with oscillator 
 radial quantum numbers up to 
 $n=M$ including 
 matrix elements of the Coulomb potential and 
 asymptotic expansion coefficients of Coulomb wave functions, with {
 $M \sim 70$} significantly exceeding the truncation boundary 
 {$N \sim 10$} of the nuclear potential 
 which will be difficult to implement in continuum extensions of {\it ab initio} approaches
 like the NCSM. 
In the method 
proposed in 
{Ref.}~\cite{Bang1999PmatrixAJ}, the Coulomb potential is cut 
at a radius~$b$ larger than the nuclear interaction radius. The conventional HORSE calculations are then performed for this modified potential 
which results are 
subsequently recalculated to obtain the phase shifts in the system with untruncated
Coulomb interaction. 
However, this approach also
encounters difficulties in many-body nuclear applications 
as the Coulomb interaction  is generated by the protons in the target which are at different
distances from the charged scattered projectile.

In this study, we propose a new approach based on the theoretical framework developed in 
{Ref.}~\cite{OKHRIMENKO1984121}. We demonstrate 
that the asymptotic three-term recurrent 
 relation (TRR) for the expansion coefficients suggested in {Ref.}~\cite{OKHRIMENKO1984121}
remains highly accurate even for small radial quantum numbers $n$. This enables the construction of a simple and 
{efficient} method for determining the scattering phase shifts. 

\section{Coulomb interaction in HORSE formalism}
\label{HORSE}

We consider the simplest single-channel case of 
scattering of two charged particles with  charges  $eZ_1$ and $eZ_2$. 
Using the partial wave expansion of the wave function, the system can be described by partial amplitudes $u_l(k, r)$ satisfying the radial Schrödinger equation with the Hamiltonian $H^l$,
\begin{equation}
    H^l u_l(k, r) = Eu_l(k, r).
    \label{shr_eq}
\end{equation}
Here, $l$ is the orbital quantum number, $r$ is the distance between the particles, $k = \sqrt{2\mu E}/\hbar$ is the momentum, $\mu$ is the reduced mass, and $E$ is the 
energy of relative motion. The functions $u_l(k, r)$ are normalized 
in such a way that the flux associated with the wave function is equal to unity.

The interaction between the particles 
$V^l = V^{Nucl,\,l} + V^{Coul}$ 
is the sum of the nuclear 
$V^{Nucl,\,l}$ and the Coulomb potentials 
$V^{Coul} = Z_1Z_2e^2/r$. 
The partial amplitude $u_l(k,r)$ in the asymptotic region $r \to \infty$ 
can be represented as a superposition 
of the regular $F_l(\eta, kr)$ and irregular $G_l(\eta, kr)$ Coulomb wave functions \cite{abramowitz1964handbook},
\begin{equation}
    u_l(k, r) 
    \xrightarrow[r \to \infty]{}
    \frac{1}{\sqrt{v}}\left[\cos{\delta_l(k)}\,F_l(\eta, kr) + \sin{\delta_l(k)}\,G_l(\eta, kr)\right],
    \label{ul_as}
\end{equation}
where $\delta_l(k)$ is the phase shift, $v = \sqrt{2E/\mu}$ is the velocity, and $\eta = \mu Z_1 Z_2 e^2 / \hbar^2 k$ is the Sommerfeld parameter.

Using the HORSE formalism, we expand the function $u_l(k, r)$ in terms of the harmonic oscillator functions $\varphi_{nl}(r)$,
\begin{equation}
    u_l(k, r) = \sum_{n=0}^{\infty}a_{nl}(k)\,\varphi_{nl}(r),
    \label{osc_series}
\end{equation}
where 
\begin{equation}
    \varphi_{nl}(r) = (-1)^n\sqrt{\frac{2 n!}{r_0\Gamma (n+l+\frac{3}{2})}}\left(\frac{r}{r_0}\right)^{\!l+1}e^{-\frac{r^2}{2r_0^2}}\,L_n^{l+\frac{1}{2}}\!\!\left(\frac{r^2}{r_0^2}\right)\!.
    \label{osc_func}
\end{equation}
Here, $\Gamma(x)$ is the gamma function~\cite{abramowitz1964handbook}, 
$L_n^\alpha(x)$ is the associated Laguerre 
polynomial~\cite{abramowitz1964handbook}, $r_0 = \sqrt{\hbar/\mu\omega}$ is the oscillator radius, and $\omega$ is the oscillator frequency.

Substituting the expansion \eqref{osc_series} into the Schrödinger equation \eqref{shr_eq}, we obtain an infinite system of linear equations for the coefficients $a_{nl}(k)$, 
\begin{equation}
    \sum_{n'=0}^{\infty}(H_{nn'}^l - \delta_{nn'}E)\,a_{n'l}(k) = 0,
    \label{matr_eq}
\end{equation}
where $H_{nn'}^l = T_{nn'}^l + V_{nn'}^l$ are 
the matrix elements of the Hamiltonian in the harmonic oscillator basis with $T_{nn'}^l$ and $V_{nn'}^l$ being 
the matrix elements of the kinetic 
$T^l$ and 
potential $V^l$ energies, 
respectively.

The kinetic energy matrix $T_{nn'}^l$ has a tridiagonal form,
\begin{equation}
    \begin{aligned}
        T_{nn'}^l&=0, \quad |n-n'|>1, \\
        T_{nn}^l&=\frac{\hbar\omega}{2}\left(2n+l+\frac{3}{2}\right),\\
        T_{n+1,\, n}^l=T_{n,\, n+1}^l&=-\frac{\hbar\omega}{2}\sqrt{(n+1)\left(n+l+\frac{3}{2}\right)},
    \end{aligned}
    \label{Tnn}
\end{equation}
and its non-zero elements $T_{nn}^l$ and $T_{n,\,n \pm 1}^l$ increase linearly with $n$ for large values of $n$ while the 
matrix elements of the short-range nuclear potential 
${V_{nn'}^{Nucl,\,l} \to 0}$ as 
$n$ and/or $n' \to \infty$. Thus, within the HORSE formalism, the potential 
$V^{Nucl,\,l}$ is replaced by 
the potential $\widetilde{V}^{Nucl,\,l}$ defined by 
a matrix in the oscillator basis
truncated at some $N$,
\begin{equation}
    \widetilde{V}_{nn'}^{Nucl,\, l}=\left\{\begin{array}{ll}
        V_{nn'}^{Nucl,\, l} & \text{for $n$ and $n' \le N$}, \\
        0 & \text{for $n$ or $n'>N$}.
    \end{array}\right .
    \label{trunc_pot}
\end{equation}
At the same time, the matrix elements of Coulomb interaction $V_{nn'}^{Coul}$ 
decrease slowly along the main diagonal and 
the diagonal matrix elements $V_{nn}^{Coul}$ should be accounted for at much larger
values of the radial quantum number~$n\gg N$~\cite{OKHRIMENKO1984121}.

Now, we consider the asymptotic region spanned by the oscillator functions $\varphi_{nl}(r)$ with $n>N$. According to the study by the Kiev group \cite{OKHRIMENKO1984121}, for large $n$, the expansion coefficients in Eq.~\eqref{matr_eq} $a_{nl}(k)\equiv a_{nl}^{as}(k)$ fit
the TRR 
\begin{equation}
    T_{n,\, n-1}^l a_{n-1,\,l}^{as}(k)+(T_{nn}^l+V_{nn}^{ad,\,l}-E)a_{nl}^{as}(k)+T_{n,\, n+1}^l a_{n+1,\, l}^{as}(k)=0, \quad n \gg 1 , 
    \label{TRR}
\end{equation}
where the additional Coulomb term
\begin{equation}
    V_{nn}^{ad,\,l} = \hbar\omega\frac{\eta kr_0}{\sqrt{4n+2l+3}}.
    \label{V_ad}
\end{equation}
The TRR 
\eqref{TRR} has two linearly independent solutions, 
 $S_{nl}(k)$ and $C_{nl}(k)$. Therefore, $a_{nl}^{as}(k)$ can be expressed as their 
superposition, 
\begin{equation}
    a_{nl}^{as}(k) = \cos{\delta_l(k)} \,S_{nl}(k) + \sin{\delta_l(k)} \,C_{nl}(k).
\end{equation}
In accordance with the wave 
function $u_l(k, r)$ asymptotic behavior~\eqref{ul_as} , 
the solutions $S_{nl}(k)$ and $C_{nl}(k)$ are defined in such a way that
\begin{align}
    \sum_{n=0}^{\infty}S_{nl}(k)\,\varphi_{nl}(r)&=\frac{1}{\sqrt{v}}F_l(\eta, kr),
    \label{Sn}
    \\
    \sum_{n=0}^{\infty}C_{nl}(k)\,\varphi_{nl}(r)&=\frac{1}{\sqrt{v}}\widetilde{G}_l(\eta, kr) \xrightarrow[r\to \infty]{}\frac{1}{\sqrt{v}}G_l(\eta, kr),
    \label{Cn}
\end{align}
where $\widetilde{G}_l(\eta, kr)$ is a 
function regular at $r=0$ 
{which fits an 
 inhomogeneous Schrödinger equation}~\cite{Zaitsev1998TrueMS}. The 
solutions $S_{nl}(k)$ and $C_{nl}(k)$ can be expressed~as 
\begin{align}
    S_{nl}(k)&=\frac{1}{\sqrt{v}}\int F_l(\eta, kr) \varphi_{nl}(r)dr,
    \label{sn_int}
    \\
    C_{nl}(k) &= \frac{1}{\sqrt{v}}\int \widetilde{G}_l(\eta, kr) \varphi_{nl}(r)dr.
    \label{cn_int}
\end{align}

The function $\varphi_{nl}(r)$ behaves asymptotically
like a delta function in the vicinity of the classical turning point $r_{turn} = \nu r_0$, where $\nu = \sqrt{4n+2l+3}$~\cite{Zaitsev1998TrueMS}:
\begin{equation}
    \varphi_{nl}(r)\xrightarrow[n\to\infty]{}\sqrt{\frac{2r_0}{\nu}}\delta(r-\nu r_0).
    \label{phi_as}
\end{equation}
Using this property, we derive from \eqref{sn_int} and \eqref{cn_int} the asymptotic expressions for the coefficients $S_{nl}(k)$ and $C_{nl}(k)$:
\begin{equation}
    S_{nl}(k) \xrightarrow[n \to \infty]{} \frac{1}{\sqrt{v}}\sqrt{\frac{2r_0}{\nu}}F_l(\eta, \nu kr_0),
    \label{sn_as}
\end{equation}
\begin{equation}
    C_{nl}(k) \xrightarrow[n \to \infty]{} \frac{1}{\sqrt{v}}\sqrt{\frac{2r_0}{\nu}}G_l(\eta, \nu kr_0).
    \label{cn_as}
\end{equation}

\section{Analysis of 
TRR for 
coefficients $S_{nl}(k)$}
\label{analysis}


First, it was important 
to establish which $n$ are large enough for the TRR~\eqref{TRR} to be valid. Suppose that for some 
{starting} value $n_{s}$ we have the coefficients $S_{n_{s}+2,\,l}(k)$ and $S_{n_{s}+1,\,l}(k)$ obtained from the asymptotic expression~\eqref{sn_as}. Using TRR, we can calculate the coefficients $S_{nl}(k)$ for ${n=0,\dots,n_s}$ and then compare them to the numerical values found by the integral \eqref{sn_int}.

\begin{figure}[t]
    \centering
    \includegraphics[width=\textwidth]{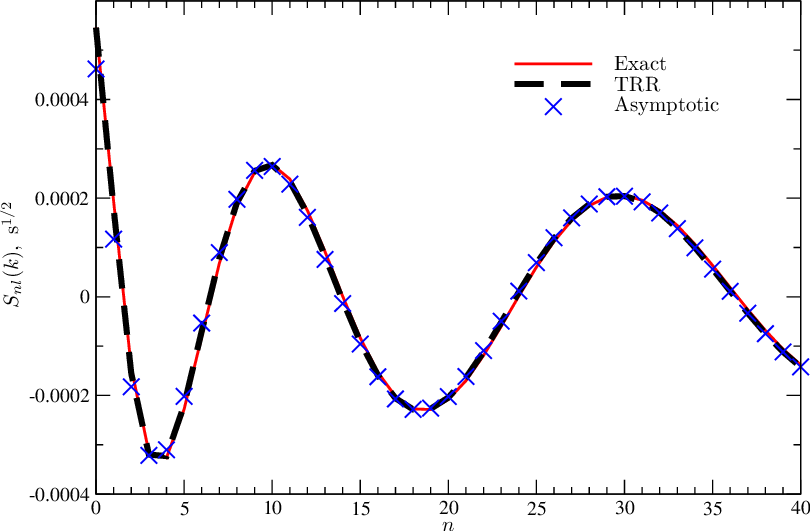}
    \caption{
Dependence of   $S_{nl}(k)$ on $n$ in   $p$-wave $p{ -} \alpha$ 
scattering ($\mu=626.4$ MeV) at $E=20$~MeV and $\hbar\omega=20$ MeV. Solid line: 
calculated by the integral \eqref{sn_int}; dashed line: 
obtained by 
TRR \eqref{TRR} starting from the asymptotic coefficients $S_{n_s+2,\,l}(k)$ and $S_{n_s+1,\,l}(k)$ with $n_s=40$; crosses: 
asymptotic values \eqref{sn_as}. 
}
    \label{fig1}
\end{figure}

The results of the calculations were 
surprising: the 
TRR allows us to reproduce the coefficients $S_{nl}(k)$ with high accuracy up to $n=0$. Moreover, the starting 
value $n_s$ 
{can be not} very large to achieve a
good convergence. An 
example of the calculations is presented 
in Fig.~\ref{fig1}. It is seen that the asymptotic expression \eqref{sn_as} for the coefficients $S_{nl}(k)$ also accurately reproduces 
the exact values even at 
small enough 
{
$n \sim 3$}. This conclusion was tested 
for different pairs of scattering particles (that is, for different $Z_1$, $Z_2$, and $\mu$) and various sets of $E$, $\hbar\omega$, and $l$ with 
similar results. However, 
for larger energies $E$, larger 
{starting} values $n_s$ were needed to obtain the same  
precision.

The 
high accuracy of the TRR~\eqref{TRR} was 
not mentioned 
and its derivation was not presented in the study of 
the Kiev group 
in Ref.~\cite{OKHRIMENKO1984121}. It is natural to assume 
that the additional Coulomb term 
$V_{nn}^{ad,\,l}$ in the TRR suggests a good approximation of 
the diagonal matrix elements of Coulomb potential $V_{nn}^{Coul,\,l}$. 
{Really, the} numerical calculations 
{demonstrated} that they differ significantly. Thus, we conclude that $V_{nn}^{ad,\,l}$ effectively represents 
{the impact of the sums which include all the Coulomb potential matrix elements} $V_{nn'}^{Coul,\,l}$, both on the main diagonal and outside it.

The analysis performed allows us to calculate similarly 
the coefficients $C_{nl}(k)$. Thus, we have developed the method to find the solutions $S_{nl}(k)$ and~$C_{nl}(k)$.

\section{Phase shifts 
}
\label{method}

Having performed a number of 
{test} calculations, we suggest the modified HORSE method for the phase shift calculations
in the case when a short-range nuclear interaction is accompanied by 
the Coulomb interaction. 
We introduce the truncated Coulomb
potential $\widetilde{V}^{Coul,\,l}$ defined by its matrix in the oscillator basis
\begin{equation}
    \widetilde{V}_{nn'}^{Coul,\, l}=\left\{\begin{array}{ll}
        V_{nn'}^{Coul,\, l} & \text{for $n$ and $n' \le N$}, \\
        0 & \text{for $n$ or $n'>N$} . 
    \end{array}\right.
    \label{trunc_pot_Coul}
\end{equation}
The convergence of the scattering observables is essentially improved if, instead of the sharp
truncated  nuclear interaction~\eqref{trunc_pot}, one makes use of a smoothly truncated 
interaction  $ {\Bbb V}^{Nucl,\,l}$ defined by its matrix in the oscillator basis~\cite{JRevai_1985}
\begin{equation}
{\Bbb V}_{nn'}^{Nucl,\,l}=\left\{\begin{array}{ll}\sigma_n V_{nn'}^{Nucl,\,l} \sigma_{n'} 
                                         & \text{for $n$ and $n' \le N$}, \\
               0 & \text{for $n$ or $n'>N$} ,
    \end{array}\right.
\label{Hungary}
\end{equation}
where
\begin{equation}
    \sigma_n = \frac{1-\exp{\{-{\left[\alpha(n-N-1)/(N+1)\right]}^2\}}}{1-\exp{\{-\alpha^2\}}}.
\end{equation}
In what follows, we use the 
smoothing parameter $\alpha=5$. 
We note that the best results are obtained when the smooth truncation~\eqref{Hungary} is used 
for the nuclear potential only while the Coulomb interaction is sharply truncated
according to Eq.~\eqref{trunc_pot_Coul}.

So, we use the Hamiltonian 
$\widetilde{H}^l = T^l+\widetilde{V}^l$ with untruncated kinetic energy~$T^l$ and effective
interaction  $\widetilde{V}^l = {\Bbb V}^{Nucl,\,l} + \widetilde{V}^{Coul,\,l}$. 
Using the technique described in 
Section~\ref{analysis}, we obtain the coefficients $S_{nl}(k)$ and $C_{nl}(k)$ for $n=N,\dots,n_s$. Having found the eigenvalues $E_{\lambda}$ and eigenvectors $\gamma_{\lambda n}$ of the truncated Hamiltonian $\widetilde{H}_{nn'}^l$ ($n,n'=0,1,\dots,N$), we use the following equation 
to calculate the phase shifts 
$\delta_l(k)$ (see \cite{Bang1999PmatrixAJ} for details):
\begin{equation}
    \tan\delta_l(k) = - \frac{S_{Nl}(k) - \mathscr{G}^{l}_{NN}S_{N+1,\,l}(k)}{C_{Nl}(k) 
    - \mathscr{G}^{l}_{NN}C_{N+1,\,l}(k)},
    \label{tan_del}
\end{equation}
where
\begin{equation}
    \mathscr{G}^{l}_{nn'} = - \sum_{\lambda=0}^N \frac{\gamma_{\lambda n}^{*}\gamma_{\lambda n'}}{E_{\lambda} - E} T^{l}_{n',\,n'+1}.
\end{equation}

\section{Results}
\label{results}

To illustrate the accuracy of the proposed approach, we use as the nuclear interaction 
the Woods--Saxon 
potential 
\begin{equation}
    V^{WS}(r)=\frac{V_0}{1+\exp{\left(\frac{r-R_0}{\alpha_0}\right)}} + (\mathbf{l} \cdot \mathbf{s})\frac{1}{r}\frac{d}{dr}\frac{V_{ls}}{1+\exp{\left(\frac{r-R_1}{\alpha_1}\right)}}
\end{equation}
with $\mathbf{l}$ and $\mathbf{s}$ denoting the orbital momentum and the spin, respectively. We compare the results of the proposed 
method with the results obtained within the approaches suggested 
{Ref.}~\cite{Bang1999PmatrixAJ} with Coulomb interaction cut at $b=7.0$~fm 
and that of Ref.~\cite{OKHRIMENKO1984121} with summation of Coulomb matrix elements
with $n\leq M=70$ as well as with the results obtained by the direct integration of the
Schr\"odinger equation by the Numerov method which we refer to as exact.

\begin{figure}[t]
    \footnotesize
    \centering
    \includegraphics[width=\textwidth]{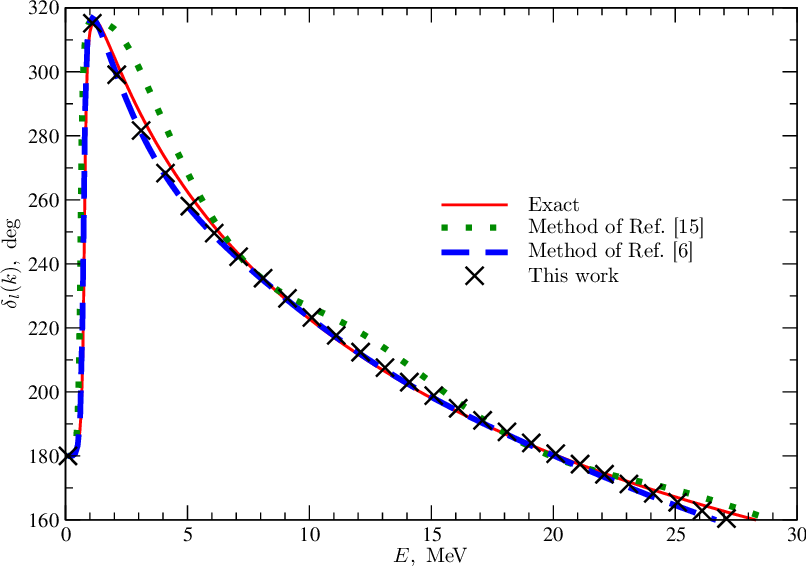}
    \caption{Phase shift $\delta_l(k)$ dependence on relative motion energy $E$ 
for $p{-}{\rm ^{15}N}$ scattering ($\mu=881.2$~MeV) in the $s$ wave obtained by different
methods.
Solid line: 
numerical integration of Schr\"odinger equation by Numerov 
method; dots: 
method suggested in Ref.~\cite{OKHRIMENKO1984121}
with $M=70$; 
dashed line: 
method suggested in Ref.~\cite{Bang1999PmatrixAJ} 
with Coulomb interaction cut at $b=7.0$~fm; 
oblique
crosses: shifts obtained by the  approach proposed here with $n_s=200$. 
In all calculations the nuclear interaction is smoothly truncated at
$N=10$ and $\hbar\omega=18$~MeV.}
    \label{fig2}
\end{figure}

{As an example, we
consider the $p{-}{\rm ^{15}N}$ 
$s$-wave scattering phase shifts. We use the 
Woods--Saxon potential parameterization suggested in Ref.~\cite{Buck}
with 
$V_0=-55.91$~MeV, $R_0=3.083$~fm, $\alpha_0=0.53$~fm, ${V_{ls}=0.9 \text{ MeV}\cdot\text{fm}^2}$, $R_1=3.083$ fm, and $\alpha_1=0.53$~fm. 
We note that the same parameterization was used in examples of calculations presented in
{Ref.}~\cite{Bang1999PmatrixAJ}.
The nuclear potential matrix is smoothly truncated at $N=10$ 
and the same $\hbar\omega=18$~MeV is used
in all 
approaches compared 
in Fig. \ref{fig2}.} It is clearly seen that the method introduced in this study provides a good convergence  
{to} the exact values and its accuracy 
is comparable with that of 
 the approach suggested in 
 {Ref.}~\cite{Bang1999PmatrixAJ}. At the same time, the Kiev group method \cite{OKHRIMENKO1984121} 
 results
 in less accurate phase shifts. 

 We 
 {examined} scattering of 
 different types of particles  at 
 different 
 angular momenta~$l$ and got 
 similar results. It was found that to obtain a better convergence for larger 
 energies $E$ it is needed only just 
 to increase $N$ and $n_s$.

\section{Conclusion}
\label{conclusion}

In this paper, we develop a 
new method for accounting for 
the long-range Coulomb interaction within the HORSE formalism. 
As in the conventional HORSE approach to scattering of uncharged  particles, 
it allows us to calculate 
the expansion 
coefficients $a_{nl}(k)$ of the wave function in infinite oscillator series 
in the asymptotic region and 
to 
find the scattering
phase shifts. 
The method demonstrates a good convergence and accuracy of the obtained phase shifts.

We believe 
that this method will be useful in applications to many-body 
and multichannel scattering problems.

\bibliographystyle{unsrt}
\bibliography{references}


\end{document}